\documentclass{aa}

\usepackage{graphicx}
\usepackage{txfonts}
\begin{document}

   \title{Non-LTE abundance corrections for late-type stars\\ from 2000Å to 3$\rm\mu m$}

   \subtitle{I. Na, Mg, and Al}


   \author{K. Lind\inst{1}
          \and 
          T. Nordlander\inst{2,3}
          \and
          A. Wehrhahn\inst{4}
          \and
          M. Montelius\inst{5}
          \and
          Y. Osorio\inst{6,7,8}
          \and
          P.~S. Barklem\inst{4}
          \and
          M. Af\c{s}ar\inst{9,10}   
          \and
          C. Sneden\inst{10}
          \and 
          C.~Kobayashi\inst{11}}

   \institute{Department of Astronomy, Stockholm University, AlbaNova University  Center, SE-106 91 Stockholm, Sweden \\
    \email{karin.lind@astro.su.se}
    \and  
    Centre of Excellence for Astrophysics in Three Dimensions (ASTRO-3D), Australia
    \and
    Research School of Astronomy and Astrophysics, Australian National University, ACT 2611, Canberra, Australia
    \and 
    Department of Physics and Astronomy, Uppsala University, Box 516, 75120 Uppsala, Sweden
    \and
    Kapteyn Astronomical Institute, University of Groningen, Landleven 12, NL-9747 AD Groningen, the Netherlands
    \and
    Isaac Newton Group, Apartado 321, E-38700 Santa Cruz de La Palma, Spain
    \and
    Instituto de Astrof\'isicade Canarias, 38205 La Laguna, Tenerife, Spain
    \and
    Departamento de Astrof\'isica, Universidad de La Laguna, 38206 La Laguna, Tenerife, Spain
    \and
    Department of Astronomy and Space Sciences, Ege University, 35100 Bornova, İzmir, Turkey
    \and
    Department of Astronomy and McDonald Observatory, The University of Texas, Austin, TX 78712, USA
    \and
    Centre for Astrophysics Research, Department of Physics, Astronomy and Mathematics, University of Hertfordshire, Hatfield, AL10 9AB, UK
             }

   \date{Received Sep. 2021; accepted }

 
  \abstract
   {It is well known that cool star atmospheres depart from local thermodynamic equilibrium (LTE). Accurate abundance determination requires taking those effects into account, but the necessary non-LTE (hereafter NLTE) calculations are often lacking.}
   {Our goal is to provide detailed estimates of NLTE effects for FGK type stars for all spectral lines from the ultraviolet to the infrared that are potentially useful as abundance diagnostics. The first paper in this series focusses on the light elements Na, Mg and Al.}
   {The code PySME is used to compute curves-of-growth for 2158 MARCS model atmospheres in the parameter range $3800<T_{\rm eff}<8000\rm K$, $0.0<\log(g)<5.5$, and $\rm-5<[Fe/H]<+0.5$. Two microturbulence values, 1 and 2\,$\rm km\,s^{-1}$, and 9 abundance points spanning $\rm-1<[X/Fe]<1$ for element X, are used to construct individual line curves-of-growth by calculating the equivalent widths of 35 Na lines, 134 Mg lines, and 34 Al lines. The lines are selected in the wavelength range between 2000Å and 3$\rm\mu m$.}
   {We demonstrate the power of the new grids with LTE and NLTE abundance analysis by means of equivalent width measurements of five benchmark stars; the Sun, Arcturus, HD84937, HD140283 and HD122563. For Na, the NLTE abundances are lower than in LTE and show markedly reduced line-to-line scatter in the metal-poor stars. For Mg, we confirm previous reports of a significant $\sim0.25$\,dex LTE ionisation imbalance in metal-poor stars that is only slightly improved in NLTE ($\sim0.18$\,dex). LTE abundances based on Mg\,II lines agree better with models of Galactic chemical evolution. For Al, NLTE calculations strongly reduce a $\sim0.6$\,dex ionisation imbalance seen in LTE for the metal-poor stars. The abundance corrections presented in this work are in good agreement with previous studies for the subset of lines that overlap, with the exception of strongly saturated lines. }
   {Consensus between different abundance diagnostics is the most powerful tool available to stellar spectroscopists to assess the accuracy of the models. Here we report that NLTE abundance analysis in general leads to improved agreement, in particular for metal-poor stars. The residual scatter is believed to be caused mainly by unresolved blends and/or poor atomic data, with the notable exception of Mg, which calls for further investigation.}


   \maketitle
%

\section{Introduction}

The chemical composition of late-type stars is important for numerous fundamental scientific questions, including but not limited to; the origin of elements, the internal physics of stars, and the formation of matter and structure from dust and planets to star clusters and entire galaxies. The amount of data collected by stellar spectroscopic surveys has virtually exploded in the last decade, as part of the "industrial revolution" of the field of Galactic archaeology \citep{2018A&ARv..26....6N,2019ARA&A..57..571J,2020ARA&A..58..205H}. Millions of optical and near-infrared spectra are now available for analysis, and the precision of inferred stellar parameters and chemical abundances is determined jointly by the quality of the spectrum, the analysis method, and the physical accuracy of the stellar models. 

Contrary to traditional spectroscopy that commonly relies on careful selection of the most reliable spectral lines, every pixel in entire spectra are now routinely analysed in the quest for higher abundance precision. This includes lines with uncertain atomic data, unknown blending properties, and departures from local thermodynamic equilibrium (LTE). The increasingly popular data-driven machine-learning techniques \citep[e.g.][]{2015ApJ...808...16N} may mitigate the known shortcomings of synthetic stellar spectra, but the model dependence still enters indirectly via the presumed known labels on which the method is trained.  

It is in general of great importance to understand the departures from LTE, the so called NLTE effects for all available abundance diagnostics. However, employing fully consistent NLTE radiative transfer for all elements in the calculation of model atmospheres and synthetic spectra is numerically far too demanding today, and as a result, a hybrid approach is commonly adopted. In the trace-element formalism, which we assume to hold also in this study, the model atmosphere is computed in LTE and the NLTE solution, i.e. the statistical equilibrium, is found independently for one element at the time. As shown in multiple studies, the impact of NLTE effects on derived chemical abundances ranges from non-existent, to moderately influential, to very strongly influential at an order of magnitude \citep[e.g.][]{2005ARA&A..43..481A,2014IAUS..298..355M,2020A&A...642A..62A}. The outcome depends strongly on the conditions of the stellar atmosphere and the spectral line of interest and NLTE effects can therefore not be predicted a priori, but must be computed for large grids of stars and lines. 

\citet{2020A&A...642A..62A} recently published large grids of NLTE departure coefficients for 13 elements. The application of these grids for GALAH spectrum analysis revealed that NLTE abundance patterns reduce the star-to-star scatter compared to LTE and effects on individual stars and lines range from strongly negative to moderately positive. That work was limited to the wavelength range and parameter range of the GALAH survey, but the departure coefficient grids have a much wider range of applicability. Here, we use the grids of \citet{2020A&A...642A..62A} to compute individual line synthetic spectra and curves-of-growth for the full FGK parameter space for as many lines as possible from the ultraviolet (UV) to the near infrared (NIR). The first paper in this series studies the astrophysically very important light elements with atomic number Z=11-13, i.e. Na, Mg, and Al. 

Sect.\,\ref{sec:obs} describes the observational data used to test the influence of the NLTE calculations. Sect.\,\ref{sec:method} describes the line selection and the calculation of LTE and NLTE synthetic spectra. Sect.\,\ref{sec:results} discusses the abundance results for the five benchmark stars, compares it to previous NLTE work and to predictions from models of Galactic chemical evolution (GCE). Sect.\,\ref{sec:conc} summarizes our conclusions.     

\section{Observations}
\label{sec:obs}

Table \ref{table:obs} lists the spectral atlases used for the five benchmark stars and their wavelength ranges. For the Sun and Arcturus, we used atlases created with the Fourier transform spectrograph (FTS) connected to the McMath/Pierce Solar telescope (Sun) and the Coude Feed/Mayall telescopes (Arcturus) at Kitt Peak National Observatory \citep{1984sfat.book.....K,1995PASP..107.1042H,2000vnia.book.....H}. For the infrared Sun, we also used the atlas created with another FTS connected to a telescope at the Institute for Astrophysics in Goettingen \citep[IAG,][]{2016A&A...587A..65R}. UV spectra were not used for the Sun and Arcturus because of the very strong line blending.\\

UV spectra for the three metal-poor stars were retrieved from the ASTRAL catalogue \citep{2013AN....334..105A} that has been created with the STIS spectrograph on the Hubble Space Telescope (HST). Optical spectra for the stars were retrieved from the archive of the Paranal Observatory Project \citep{2003Msngr.114...10B}, collected with the UVES spectrograph on the VLT.  For two stars, HD140283 and HD122563, we also include infrared spectra observed with the IGRINS instrument on the Harlan J.~Smith Telescope at McDonald Observatory \citep{2016ApJ...819..103A}.     

\begin{table*}
\caption{Overview of the spectral atlases used, listing instrument/telescope and wavelength region.}              
\label{table:obs}      
\centering                                      
\begin{tabular}{l c c c c c c}          
\hline\hline                        
Star & \multicolumn{2}{c}{UV} & \multicolumn{2}{c}{VIS}  & \multicolumn{2}{c}{IR}  \\    
\hline                                   
    Sun & \multicolumn{2}{c}{-}  &  FTS/McMath-Pierce$^{(a)}$ & 3726-9300Å & FTS/AIG$^{(b)}$ & 4047-22994Å\\      
    Arcturus & \multicolumn{2}{c}{-}  &  FTS/Coude Feed$^{(c)}$ & 3726-9300Å & FTS/Mayall$^{(d)}$ & 9127-53547Å\\      %
    HD84937 & STIS/HST$^{(e)}$ & 2276-3116Å  &  UVES/VLT$^{(f)}$ & 3040-10400Å & \multicolumn{2}{c}{-}  \\  
    HD140283 & STIS/HST$^{(e)}$ & 1937-3159Å  &  UVES/VLT$^{(f)}$ & 3040-10400Å & IGRINS/Smith$^{(g)}$ & 14785-24780Å \\  
    HD122563 & STIS/HST$^{(e)}$ & 2276-3116Å  &  UVES/VLT$^{(f}$ & 3040-10400Å & IGRINS/Smith$^{(g)}$ & 14785-24780Å \\  
\hline                                             
    \multicolumn{7}{l}{$^{(a)}$ \citet{1984sfat.book.....K}, used $<$9000Å \ \ $^{(b)}$ \citet{2016A&A...587A..65R}, used $>$9000Å \ \ $^{(c)}$ \citet{2000vnia.book.....H}, used $<$9200Å}\\
    \multicolumn{7}{l}{$^{(d)}$ \citet{1995PASP..107.1042H}, used $>$9200Å \ \ $^{(e)}$ \citet{2013AN....334..105A}, used $<$3050Å \ \ $^{(f)}$ \citet{2003Msngr.114...10B}, used $>$3050Å}\\ 
    \multicolumn{7}{l}{$^{(g)}$ \citet{2016ApJ...819..103A}}
\end{tabular}
\end{table*}

\section{Method}
\label{sec:method}

\subsection{Line selection}
The lines used for abundance determination were only selected based on the prospects of accurately measuring their equivalent widths in (some part of) the covered parameter space. Lines with very close blends by other species, either predicted as being present in the line list or visually evident from asymmetry in their line profile, were discarded.  
The line atomic data were retrieved from VALD \footnote{\url{http://vald.astro.uu.se}} using "extract all" requests with default settings and including hyperfine splitting. The quality of available atomic data was not used as a selection criterion, and we advise the reader to take this into consideration when using the grids for abundance measurements. 

The continuum was placed individually by eye for each line. For the Sun and Arcturus, the original normalisation of the spectral atlases was trusted to be accurate to within $\sim0.5\%$. The equivalent widths were derived with Gaussian line fits using the IDL SolarSoft library\footnote{\url{http://www.lmsal.com/solarsoft/}}. Multiple Gaussians were used to de-blend partially blended lines. For strong lines, we model the intrinsic Voigt profile using overlapping Gaussians of different depth and width. The observed equivalent widths were computed by integrating the fitted line profiles and range from $1-1000$m\AA\,, which is a much more extended range than usually used for precision spectroscopy. The choice was made to demonstrate the influence of NLTE for as many lines as possible, but the average stellar abundances shown in the paper are based on a more restricted selection of lines. Good agreement was found when comparing to the equivalent widths found by direct integration of observed line profiles of the most unblended lines. 

The equivalent width error was estimated as the sum of three components; the formal fitting uncertainty as determined with the \citet{1988IAUS..132..345C} formula\footnote{$1.5/(S/N)\times\sqrt(\lambda\delta\lambda/R)$, where S/N is the average flux error, $\lambda$ the wavelength, $\delta\lambda$ the pixel size, and $R$ the spectral resolution}, an additional $0.5\%$ contribution stemming from the uncertainty in the continuum placement, and finally an error due to the blending fraction. The latter was estimated as 25$\%$ of the difference in equivalent width found by directly integrating the observed and synthetic spectra in regions where the synthetic flux is deeper than 0.1$\%$. It should be clear to the reader that this is an empirical approach that mainly is useful to compare relative sizes of errors between lines. We omit errors from other sources, such as atomic data.      

\subsection{Synthetic spectra}
We compute LTE and NLTE synthetic spectra using the code PySME v0.4.142 \citep[][and Wehrhahn et al. in prep.]{2021csss.confE...1W}\footnote{\url{sme.astro.uu.se}}, a derivative of the classical and well-used Spectroscopy Made Easy \citep{2017A&A...597A..16P} with the IDL-libraries re-implemented in python. A grid of 2158 MARCS models in the parameter range $3800<T_{\rm eff}<8000\rm K$, $0.0<\log(g)<5.5$, and $\rm-5<[Fe/H]<+0.5$, was used assuming two values of microturbulence, 1 and 2\,$\rm km\,s^{-1}$. The abundance $\rm-1<[X/Fe]<1$ was varied in steps of 0.25\,dex, resulting in 9 abundance points for each element X. For each line, 38844 LTE and NLTE synthetic spectra were thus computed. In total, approximately 200,000 CPU hours were used for the calculations. 

The spectra were computed individually, on a wavelength grid spanning $\pm100\AA$ centred on each line. Wavelength points with normalized synthetic flux deeper than 0.99999 were saved and used to compute equivalent widths by integrating the flux profiles. NLTE abundance corrections were subsequently computed by interpolating the LTE and NLTE curves-of-growth to observed equivalent widths.    

To synthesize lines in NLTE, PySME takes as input pre-computed grids of departure coefficients, i.e., ratios of NLTE/LTE level populations as a function of atmospheric depth. We used version 3 of the grids\footnote{\url{https://zenodo.org/record/3982506#.YGG_2S9yrx5}} recently published by \citet{2020A&A...642A..62A} that were computed with the NLTE code Balder \citep{2018A&A...615A.139A}, and make sure that PySME uses the same grid of  atmospheric models as Balder. The atomic models used by \citet{2020A&A...642A..62A} to compute departure coefficients were originally constructed for Na by \citet{2011A&A...528A.103L}, Mg by \citet{2015A&A...579A..53O}, and Al by \citet{2017A&A...607A..75N}.   

The solar abundance scale adopted as reference in this paper \citep{2007SSRv..130..105G} is the same as used in the MARCS models and the departure coefficient grid, specifically $A\rm(Na)=6.17$, $A\rm(Mg)=7.53$, $A\rm(Al)=6.37$, and $A\rm(Fe)=7.45$.

\subsection{Stellar parameters}
The stellar parameters adopted for the five benchmark stars are summarized in Table\,\ref{table:obs}. The effective temperatures for stars other than the Sun have been computed by bolometric fluxes and limb-darkened angular diameters. In the case of Arcturus, HD140283 and HD122563, the diameters have been directly measured by interferometry, while for HD84937, surface-brightness relations were used to obtain indirect diameters by \citet{2015A&A...582A..49H}. Surface gravities for the stars other than the Sun have been computed with the fundamental relation to mass and radius.   

Metallicities and microturbulence values that can be found in the literature are based on different stellar parameters and/or different radiative transfer codes and/or different modelling assumptions (1D/3D, LTE/NLTE). To ensure homogeneity, we therefore redetermine the metallicity, [Fe/H], and microturbulence values, $\xi$, for all the stars by measuring the strength of 11-14 Fe\,II lines and enforcing a flat trend in LTE abundance with equivalent width. The lines were selected based on the recommendations of \citet{2021A&A...645A.106H} and are summarized in Table\,\ref{table:Fe}. The atomic data can be found in \citeauthor{2021A&A...645A.106H}. We note that the metallicity scale for the metal-poor stars is in agreement within 0.03dex with the recent study by \citet{2016MNRAS.463.1518A}, when limiting the comparison to Fe\,II lines modelled in 1D LTE.

\begin{table}
\caption{Adopted stellar parameters}              
\label{table:para}      
\centering                                      
\begin{tabular}{l c c c c }          
\hline\hline                        
Star & $T_{\rm eff}$ & $\log(g)$ & [Fe/H] & $\xi$ \\    
 & [K] & & & [km/s] \\
\hline                                   
    Sun &       5772$^{(a)}$ & 4.44$^{(a)}$ &  0.0          & 0.9$^{(d)}$ \\      
    Arcturus &  4286$^{(b)}$ & 1.64$^{(b)}$ & -0.55$^{(d)}$ & 1.3$^{(d)}$\\      
    HD84937 &   6356$^{(b)}$ & 4.06$^{(b)}$ & -2.06$^{(d)}$ & 1.2$^{(d)}$\\  
    HD140283 &  5792$^{(c)}$ & 3.65$^{(c)}$ & -2.38$^{(d)}$ & 1.3$^{(d)}$\\  
    HD122563 &  4636$^{(c)}$ & 1.40$^{(c)}$ & -2.48$^{(d)}$ & 1.8$^{(d)}$\\  
\hline                                             
    \multicolumn{2}{l}{$^{(a)}$ \citet{2016AJ....152...41P}} & \multicolumn{3}{l}{$^{(c)}$ \citet{2020A&A...640A..25K}}\\
    \multicolumn{2}{l}{$^{(b)}$ \citet{2015A&A...582A..49H}} &     \multicolumn{3}{l}{$^{(d)}$ This work}\\
\end{tabular}
\end{table}

\section{Results and discussion}
\label{sec:results}

\subsection{Na}
Departures from LTE for Na\,I have been studied extensively, starting with pioneering efforts half a century ago \citep{1969ApJ...156..695A,1975A&A....38..289G}. More recent references can be found in \citet{2011A&A...528A.103L} and \citet{2014AstL...40..406A}. The dominant NLTE effect in Na\,I is darkening of the cores of strong lines, caused by photon suction. NLTE corrections based on equivalent-width measurements are negative and can reach more than -0.5\,dex at full saturation. Due to the sparsity of lines of this alkali in late-type spectra, saturated lines are sometime the only available diagnostic. One extreme example is the use of the Na\,D lines in metal-poor horizontal branch stars, with NLTE corrections as large as $-0.9$\,dex \citep{2011ApJ...730L..16M}. 

Grids of NLTE corrections for Na\,I lines have been published by several authors, all solving the restricted NLTE problem for trace elements in 1D LTE stellar atmospheres:
\begin{itemize}
    \item \citet{2000ARep...44..790M}, for 14 optical/NIR lines $<1.3\rm\mu m$
    \item \citet{2003ChJAA...3..316T}, for 8 optical lines
    \item \citet{2004A&A...423..683S}, for 8 optical lines
    \item \citet{2007A&A...464.1081A}, for the Na\,D lines      
    \item \citet{2011A&A...528A.103L}, for 11 optical/NIR lines $<1.1\rm\mu m$
    \item This study includes 35 Na\,I UV-NIR lines.
\end{itemize}

The optical wavelength band has traditionally dominated studies of Galactic archaeology, but with the advent of high-resolution multi-object spectrographs like APOGEE \citep{2017AJ....154...94M}, the NIR is attracting more and more attention. To our knowledge, the first work to study NLTE effects on the $1.6\rm\mu m$ Na\,I lines in the H band is \citet{2020A&A...637A..80O}, but as their method is based on spectrum synthesis, they did not publish a grid of NLTE corrections. In this study, we compute and make available corrections for 35 Na\,I lines between $\rm 3302Å$ and $2.59\rm\mu m$, thus extending the availability of such data both into the UV and further into the NIR. All oscillator strength have high quality, with A-B ratings in NIST. New broadening data due to hydrogen collisions were computed for three NIR lines around 2.2$\mu m$ using the Anstee, Barklem \& O'Mara theory \citep{1995MNRAS.276..859A,1997MNRAS.290..102B}; in the case of the 2.14 micron line the upper state is a Rydberg state and an estimate of the broadening was made following \citet{2015A&A...579A..53O} \citep[see also][]{1995JPhB...28.3147H}.

\begin{figure*}
    \centering
    \includegraphics[width=\textwidth,viewport=0cm 0cm 21cm 26cm, clip=true]{"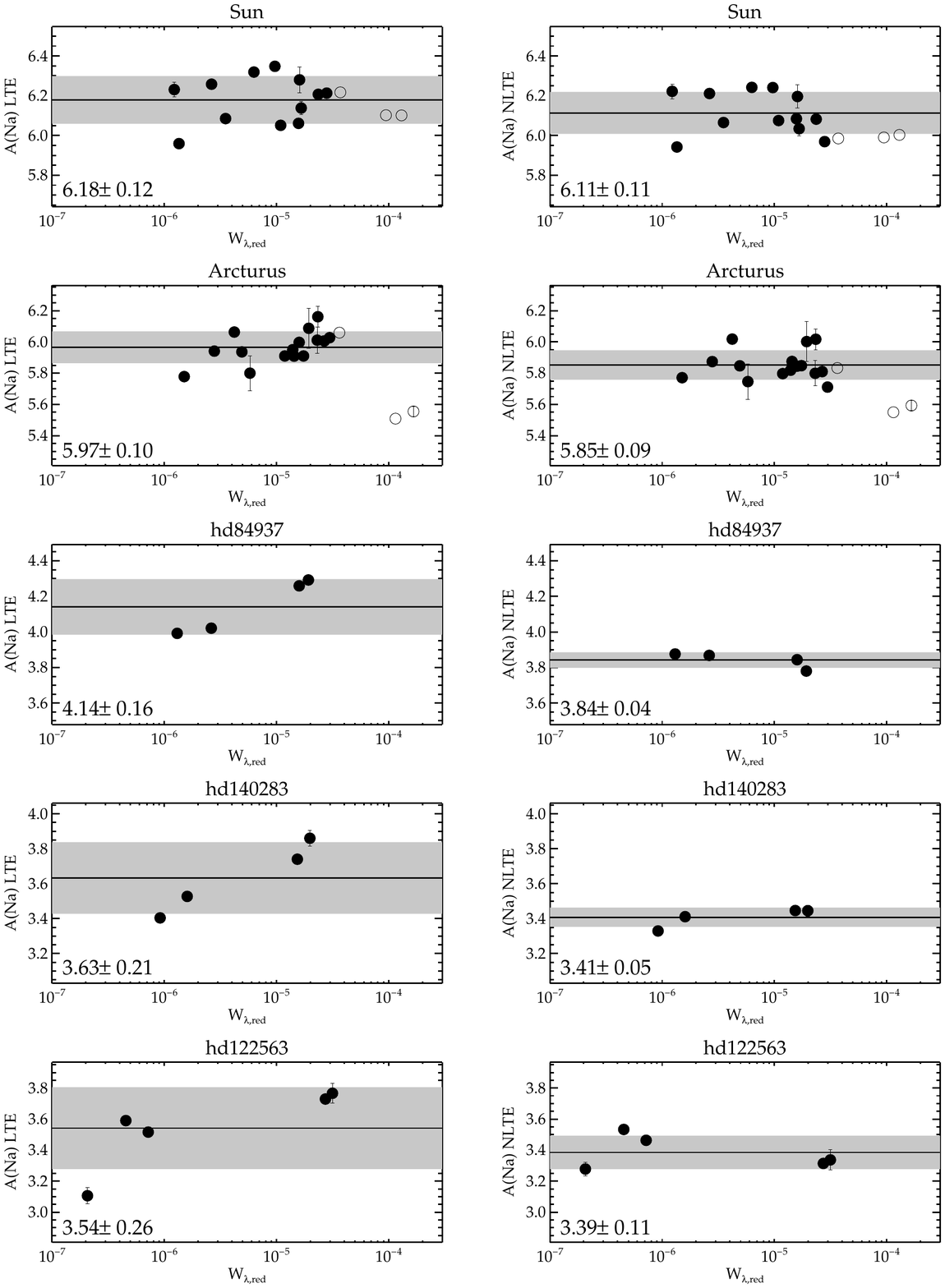"}
    \caption{Na abundances derived from neutral lines with $W_{\lambda,red}<3\times10^{-5}$ (black bullets) and $W_{\lambda,red}>3\times10^{-5}$ (open bullets) in LTE (left) and NLTE (right). The horizontal lines and grey shaded regions correspond to the mean and one sigma standard deviation of the lines.}
    \label{fig:Na}
\end{figure*}

Fig. \ref{fig:Na} shows the Na LTE and NLTE abundances determined for the five benchmark stars in our sample. The abundance and line data can be found in Table\,A\ref{table:Na}. For the Sun and Arcturus, 13-14 lines are available in the optical ($\rm<1.1\mu m$) and 2-4 lines around $\rm 2\mu m$. NLTE effects are moderate, lowering the average abundance by -0.1\,dex. The line-to-line scatter in the Sun and Arcturus is slightly reduced.

In the metal-poor stars, only 3-5 lines are strong enough to be measured. The NLTE abundances are lower by $-0.2$\,dex on average, and a drastically reduced scatter is seen for all stars. This is primarily driven by the strong negative NLTE abundance correction for the Na\,D lines. Overall, acceptable agreement is seen across UV, optical, and NIR wavelength ranges, when available.

\begin{figure*}
    \centering
    \includegraphics[width=\textwidth,viewport=4cm 13cm 31cm 21cm, clip=true]{"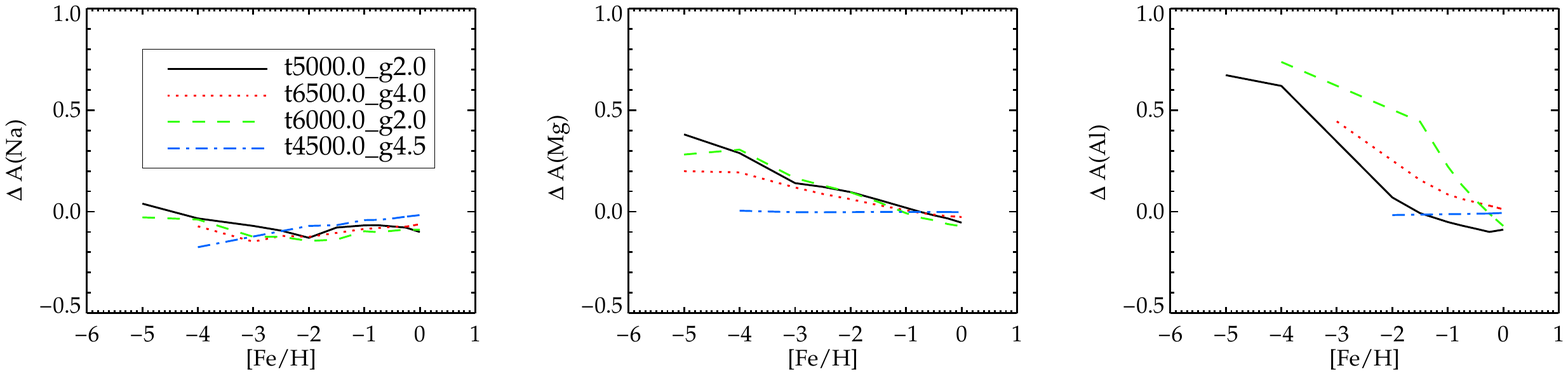"}
    \caption{The figure illustrates the average NLTE effect for optical unsaturated lines ($3000-10,000$\AA\ and $W_{\lambda,red}<10^{-5}$) as a function of metallicity. Four different stellar types are shown with stellar parameters as indicated by the legend. Only those data points are retained that have $<0.1$\,dex standard deviation among the spectral lines used to form the mean.}
    \label{fig:cases}
\end{figure*}

Fig.\,\ref{fig:cases} illustrate how the typical NLTE effect for unsaturated lines vary as a function of stellar parameters. Four different stellar types are shown, one red giant, one horizontal branch star, one cool dwarf and a turn-off star. We see that NLTE corrections for unsaturated lines are typically small and negative, $-0.1$\,dex to $-0.2$\,dex, everywhere in the displayed parameter space. Only at the very lowest metallicities, a slight positive upturn is seen.

\subsection{Mg}
Mg is typically twenty times more abundant than its two neighbours in the periodic table, and many more lines are also available as abundance diagnostics in late-type stars. Lines of both neutral and singly ionized Mg can be measured down to low metallicity. The first NLTE investigation again dates back to \citet{1969ApJ...156..695A} and many have followed since then. Several studies have determined NLTE line formation for a specific sample of stars, e.g. \citet{2010A&A...509A..88A}, \citet{2017ApJ...847...16B}, and \citet{2019MNRAS.485.3527S}, but only a handful of studies have provided NLTE abundance corrections in tabulated form: 

\begin{itemize}
    \item \citet{2000ARep...44..530S} for five optical lines
    \item \citet{2011MNRAS.418..863M} for 16 optical/NIR lines $\rm<1.2\mu m$
    \item \citet{2015ApJ...804..113B} for two NIR lines
    \item \citet{2016A&A...586A.120O} for 19 optical/NIR lines $\rm<2.1\mu m$ 
    \item This study includes 134 Mg\,I UV-NIR lines.
\end{itemize}

For Mg\,I, NLTE effects are typically small and can be both positive and negative, depending on the competing effects of overionization and photon suction. Because the model atom \citep{2015A&A...579A..53O} lacks collisions between hydrogen atoms and Mg\,II, we suspect that NLTE effects are overestimated for Mg\,II and assume that LTE is a better approximation. 

\begin{figure*}
    \centering
    \includegraphics[width=\textwidth,viewport=0cm 0cm 21cm 26cm, clip=true]{"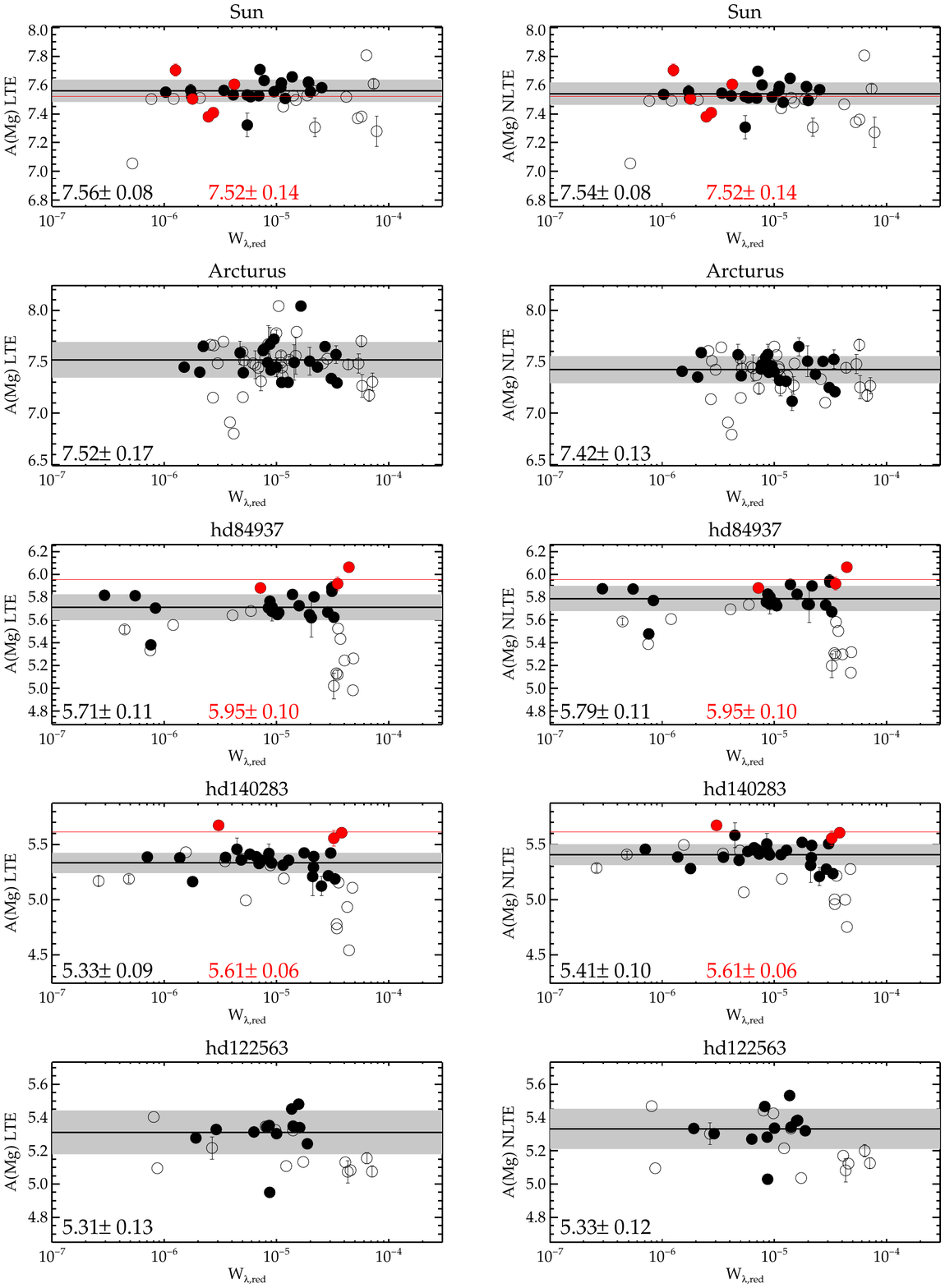"}
    \caption{Mg abundances derived from neutral lines with $W_{\lambda,red}<3\times10^{-5}$ and high-quality $gf$-values (black bullets), from neutral lines with $W_{\lambda,red}>3\times10^{-5}$ and/or low-quality $gf$-values (open bullets), and from singly ionized lines (red bullets) in LTE (left) and NLTE (right). The black horizontal lines and grey shaded regions correspond to the mean and one sigma standard deviation of the lines marked with filled black bullets. The red horizontal lines correspond to the mean of the singly ionized lines.}
    \label{fig:Mg}
\end{figure*}

Many Mg\,I lines have oscillator strengths with large uncertainties in the VALD3 database, with corresponding entries in NIST flagged as C ($\leq25\%$) or lower. The transition probabilities have been much improved with the combined experimental and theoretical study by \citet{2017A&A...598A.102P}, and we used this reference when possible. In Fig. \ref{fig:Mg}, we show Mg\,I-based abundances determined with \citeauthor{2017A&A...598A.102P} values with filled black bullets and other Mg\,I lines with open bullets. Red bullets mark Mg\,II lines. The line-by-line abundances are listed in Table\,A.\ref{table:Mg}.

Between 26 and 55 lines are used for abundance determination of the five benchmark stars, which is significantly more than previous studies. The average NLTE effect for Mg\,I lines is $-0.02$\,dex in the Sun, $-0.10$\,dex in Arcturus and $+0.08$\,dex for the metal-poor turn-off and subgiant. As the plots clearly show, the line-to-line scatter is smaller when only considering lines that have high-quality transition probabilities. The line-to-line scatter for Mg\,I lines is not significantly affected by NLTE. In both LTE and NLTE there is a tendency for saturated lines to give lower abundance. In particular, the saturated lines <3000\AA\ give abundances that are significantly lower than the average. The mean abundances and standard deviations shown in Fig.\,\ref{fig:Mg} are based on weak to moderately strong lines with good oscillator strengths, as indicated in the figure caption.    

Mg\,II lines are not commonly used for abundance analysis in the literature, and it is particularly important to stress the significant ionization imbalance, of order $\sim0.25$\,dex in LTE and only mildly reduced to $\sim0.18$\,dex in NLTE, found for HD84937 and HD140283. This confirms the results of \citet{2018ApJ...866..153A} who reported a clear discrepancy between abundances based on Mg\,II 4481\AA\ and the average of Mg\,I lines for the same two stars. Here, we add two bluer Mg\,II lines that support the high abundance found with 4481\AA\ , and hence undetected blends in the latter is probably not the underlying cause. At this point, we may only speculate along the same lines as \citeauthor{2018ApJ...866..153A} in that 3D effects may be to blame for the ionization imbalance, and we plan to tackle Mg with full 3D NLTE modelling in a future study. However, according to \citet{2017ApJ...847...15B}, NLTE abundances computed based on Mg\,I lines using so called $\rm<3D>$-models, which are spatial and temporal averages of full 3D simulations, are not substantially higher than 1D-based abundances. We cannot rule out that our atomic model or modelling assumptions underestimates the NLTE effects in Mg\,I in both 1D and 3D, perhaps because the trace element method neglects possible feedback effects from other elements \citep{2020A&A...637A..80O}. 

Looking at a more extended parameter space, Fig.\,\ref{fig:cases} shows that our predicted Mg abundance corrections for unsaturated lines show a simple trend that is increasingly positive toward lower metallicites. At the lowest metallicities, giants are predicted to depart from LTE by $\sim0.3$\,dex.

%

\subsection{Al}

NLTE spectroscopic analysis for Al\,I was mainly pioneered by \citet{1996A&A...307..961B,1997A&A...325.1088B}, highlighting the strong overionization in metal-poor stars. Later important work include, e.g., \citet{2004A&A...413.1045G,2006A&A...451.1065G} and \citet{2008A&A...481..481A}, which however did not compute grids of abundance corrections to use for quantitative spectroscopy. The first such grids were published by:

\begin{itemize}
    \item \citet{2012AstBu..67..294M} for 9 optical lines
    \item \citet{2016AstL...42..366M} for 6 optical lines
    \item \citet{2017A&A...607A..75N} 26 UV-NIR lines
    \item This study includes 43 Al\,I + 3 Al\,II UV-NIR lines 
\end{itemize}

The number of Al lines available for abundance analysis in the five benchmark stars is similar to Na, i.e., around 18 in total at high metallicity, but only a handful at low metallicity. The main reason why Al is nevertheless a less well studied element is that the detectable lines at low metallicity fall below 4000Å or in the H band, while Na has strong lines in the optical. The NLTE effects are also similar at high metallicity, with photon suction causing increasingly negative corrections as the lines become saturated. At low metallicity, however, the overionization of neutral Al is much stronger than for both Na and Mg and the Al\,I NLTE abundance corrections are therefore positive, of order $+0.4$\,dex in the two unevolved stars. 

\begin{figure*}
    \centering
    \includegraphics[width=\textwidth,viewport=0cm 0cm 21cm 26cm, clip=true]{"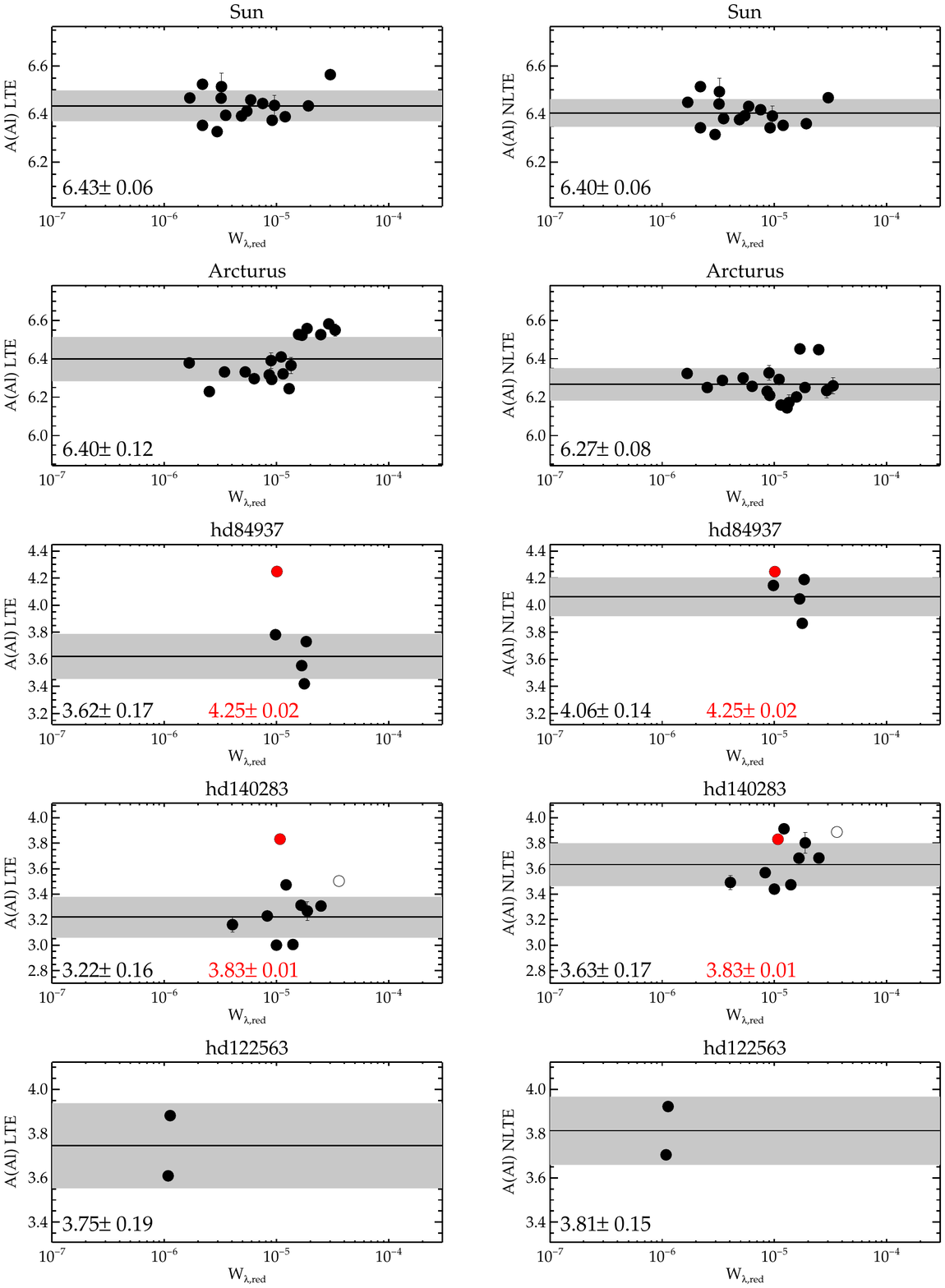"}
    \caption{Al abundances derived from neutral lines with $W_{\lambda,red}<3\times10^{-5}$ and high-quality $gf$-values (black bullets), from neutral lines with $W_{\lambda,red}>3\times10^{-5}$ and/or low-quality $gf$-values (open bullets), and from singly ionized lines (red bullets) in LTE (left) and NLTE (right). The black horizontal lines and grey shaded regions correspond to the mean and one sigma standard deviation of the lines marked with filled black bullets.}
    \label{fig:Al}
\end{figure*}

As shown in Fig.\ref{fig:Al} and Table\,A\ref{table:Al}, the line-to-line scatter for neutral lines is not strongly affected overall, but we note that for Arcturus, NLTE analysis brings the strong NIR lines into agreement with the optical lines. For HD140283 and HD84937, we can investigate the ionization balance using Al\,II 2669\AA\, which forms close to LTE and has a high-quality $gf$-value from \citet{Tr_bert_1999}. For the neutral lines, we adopted the theoretical gf-values available in TOPbase \citep{1995JPhB...28.3485M}. We find that the abundance determined from the singly ionized line is higher than the neutral lines by $0.6$\,dex in LTE and $0.2$\,dex in NLTE. Given the fewer number of lines and their scatter, we consider the ionization imbalance in NLTE for Al less alarming than for Mg. The ionization balance of Al was first investigated by \citet{2021ApJ...912..119R} for a sample of 11 metal-poor stars, finding a 0.4-0.9\,dex discrepancy between the ionization stages in LTE. They further noted that the NLTE corrections provided by \citet{2017A&A...607A..75N} resolved the imbalance.   

Fig.\,\ref{fig:cases} emphasizes further how important NLTE analysis is for Al abundances of metal-poor stars, with corrections of up to $0.8$\,dex for giants in the metal-poor end of the studied grid. Cool dwarf stars show a different behaviour, with negative corrections in the metallicity range where unsaturated optical lines are detectable. 

\subsection{Comparison to previous work}

\begin{figure*}
    \centering
    \includegraphics[width=\textwidth,viewport=0cm 0cm 21cm 26cm, clip=true]{"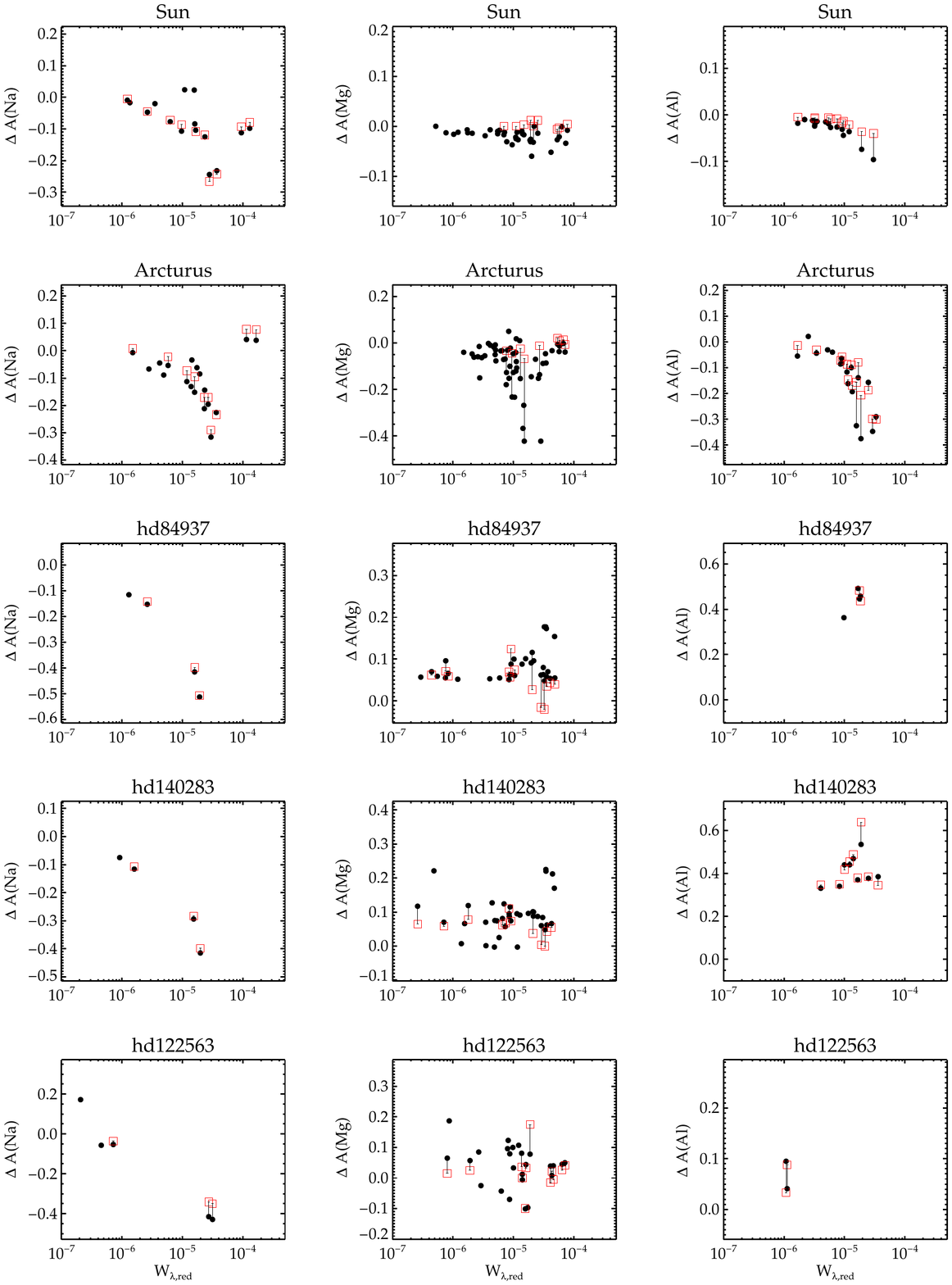"}
    \caption{The new NLTE corrections compared to previous work. Black bullets represent the results of this study, red open squares the results of \citet{2011A&A...528A.103L} for Na, \citet{2016A&A...586A.120O} for Mg, and \citet{2017A&A...607A..75N} for Al. Vertical lines connects lines in common.}
    \label{fig:delt}
\end{figure*}

In Sect. 4.1-4.3 we listed several literature studies that present NLTE abundance corrections for Na, Mg, and Al for a grid of 1D model atmospheres. As a whole, these present a variety of different atmospheric and atomic models, as well as different codes for statistical equilibrium and radiative transfer calculations. In common for all three elements, is that quantum mechanical data for collisions with neutral hydrogen have become available in the last decade \citep{2010PhRvA..81c2706B,2013A&A...560A..60B,2010A&A...519A..20B,2012A&A...541A..80B}, thereby removing a notorious and large source of uncertainty in the atomic model. The impact of the new data on stellar abundances with respect to the previously adopted Drawin formula \citep{1968ZPhy..211..404D,1969ZPhy..225..483D} have been extensively described \citep[e.g.][]{2013A&A...550A..28M,2016AstL...42..366M,2018A&A...618A.141E}. Here, we limit the comparison to work that use, with minor modifications, the same atomic model and MARCS atmospheric models as was used by \citet{2020A&A...642A..62A} to compute the departure coefficients adopted in this work. The comparison therefore mainly highlights differences caused by choice of code to solve the statistical equilibrium and compute the synthetic spectrum. All previous work shown in Fig.\,\ref{fig:delt} used MULTI2.3 \citep{1992ASPC...26..499C} for both purposes, while we have used Balder to establish the statistical equilibrium and PySME to compute the curves-or-growth.    

In Fig.\ref{fig:delt} we compare our results to \citet{2011A&A...528A.103L} for Na, \citet{2016A&A...586A.120O} for Mg, and \citet{2017A&A...607A..75N} for Al. The agreement for the majority of lines is good, but there are several notable exceptions, in particular for saturated lines with reduced equivalent widths in the range $W_{\lambda,red}=(1-3)\times10^{-5}$. For such lines, our new corrections can be significantly different, and usually more negative. The effect is strongest for the two giants in our sample. To attempt to trace the origin of these differences, we have investigated the curves-of-growth in some more detail. 

\begin{figure*}
    \centering
    \includegraphics[width=\textwidth,viewport=4cm 13cm 31cm 21cm, clip=true]{"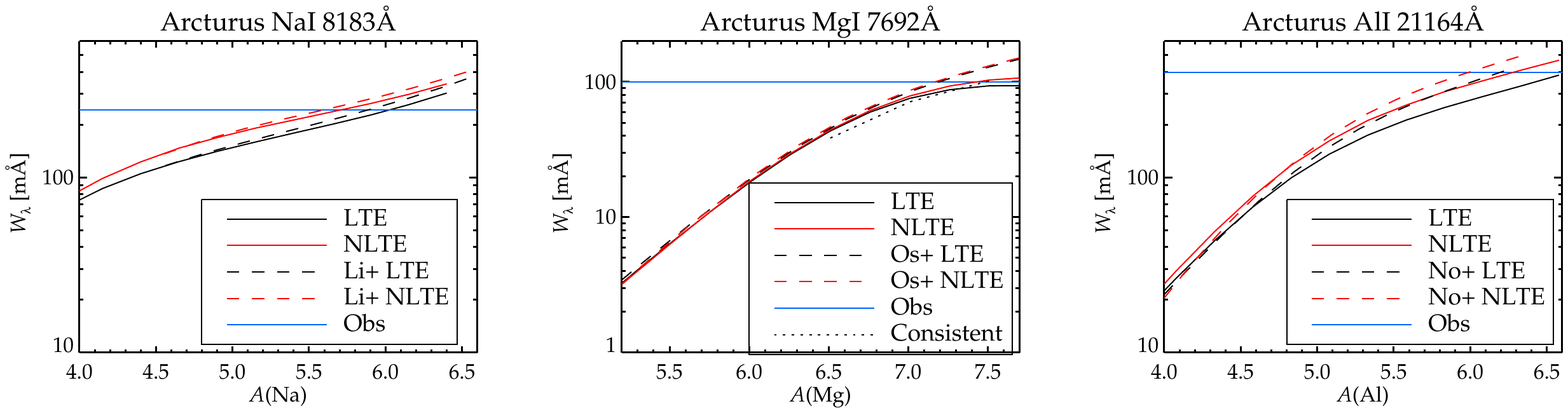"}
    \caption{Examples of LTE (black) and NLTE (red) curves-of-growth for spectral lines of three different elements in Arcturus. Solid lines represent the results of this work, dashed lines the results of \citet{2011A&A...528A.103L} for Na, \citet{2016A&A...586A.120O} for Mg, and \citet{2017A&A...607A..75N} for Al. The black dotted line in the middle panel has been computed with model atmospheres that have consistent $\rm[\alpha/Fe]$-ratios as explained in the text.}
    \label{fig:cogs}
\end{figure*}

Fig.\ref{fig:cogs} illustrates how the line strength grows in LTE and NLTE for three lines that are saturated in Arcturus; Na\,8183\AA\,, Mg\,7692\AA\,, and Al\,21164\AA\,.  Our new curves-of-growth computed with Balder/PySME are compared to ones computed with MULTI2.3. We see that in all three cases that the agreement between LTE curves and NLTE curves at a given line strength is similar, i.e., the new results deviate in unison from the old. This leads us to conclude that the differences do not originate in different solutions to the statistical equilibrium, but from calculation of LTE and NLTE line and continuous opacity by MULTI2.3 and PySME.

We note the flattening of the curves for Mg that is evident at high abundance, where the equivalent width appears insensitive to increased abundance. In fact, this is a behaviour seen for all Mg lines, and at $\rm[Mg/Fe]>1$, the curves can even have a negative slope. This counter-intuitive behaviour can be explained by the importance of Mg as electron donor. As the abundance increases, the continuous opacity which is dominated by H$^-$ also increases, reducing spectral line strengths. If the atmospheric models had been recomputed consistently with the adopted chemical composition, the trends may well look different. The dotted line marked "Consistent" in the Mg panel illustrates this by showing a smaller part of the LTE curve-of-growth for which we have used alpha-negative ($\rm[\alpha/Fe]=-0.4$), alpha-poor ($\rm[\alpha/Fe]=0.0$) and alpha-enhanced MARCS models ($\rm[\alpha/Fe]=0.4$) instead of the standard composition ($\rm[\alpha/Fe]=0.2$ at $\rm[Fe/H]=-0.5$). This curve crosses the reference and inconsistent LTE curve-of-growth (black solid) at $\rm[\alpha/Fe]=0.2$ as expected, but does not show the same flattening at higher abundance. We therefore advise caution in the use of the grids at the upper and lower end of the abundance range.                


\begin{figure*}
    \centering
    \includegraphics[width=\textwidth,viewport=4cm 13cm 31cm 21cm, clip=true]{"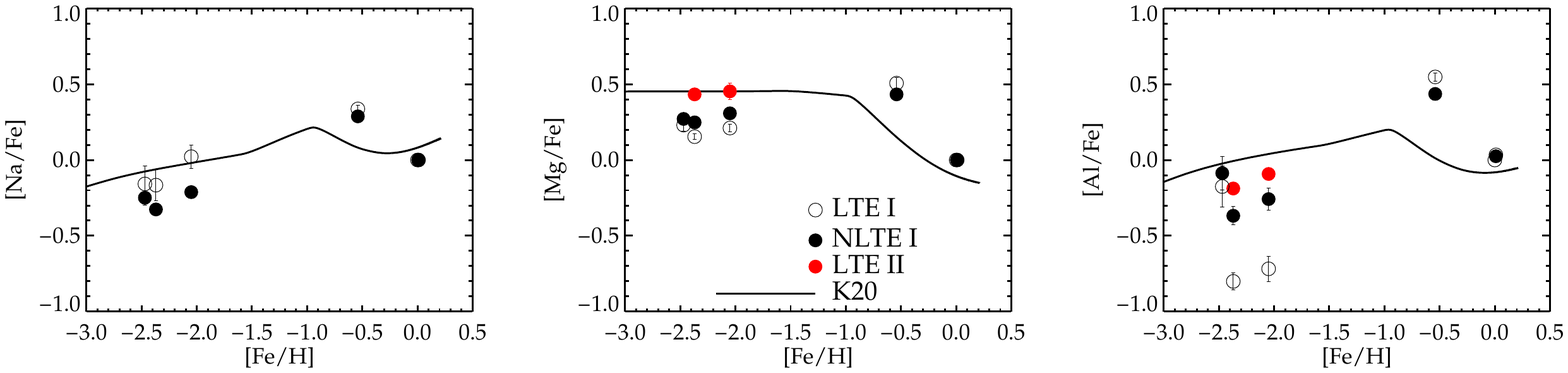"}
    \caption{Comparison of the abundances found for the benchmark stars analysed in this paper (bullets) to the Galactic chemical evolution model of \citet{2020ApJ...900..179K}. The Sun is marked at the origin. }
    \label{fig:gce}
\end{figure*}

\subsection{Comparison to Galactic chemical evolution models}

All three elements studied here are primarily forged in massive stars during hydrostatic carbon burning and returned to the interstellar medium by core-collapse supernovae (SN) and hypernovae (HN) \citep{2013ARA&A..51..457N,2019Sci...363..474J}. The $\alpha$-element Mg is predicted to form a plateau value in [Mg/Fe] at low metallicity and decreasing ratio when supernovae of type 1a (SNIa) start contributing with relatively more Fe. The metallicity at which this decrease occurs is environment dependent, because the star formation history depicts how much Fe had time to form via SN and HN prior to the delayed onset of SNIa. In the Milky Way, the "knee" in the $\rm[\alpha/Fe]$ is located at $\rm[Fe/H]=-1$. 

Na and Al yields, however, depend also on the available neutron excess and thus on the metallicity of the massive star progenitor. The [Na/Fe] and [Al/Fe] ratios are therefore predicted to increase with increasing metallicity, until SNIa start contributing and reduce the ratios. For completeness, it should be mentioned that all three elements have minor contributions from  the NeNa and MgAl nucleosynthesis channels that coexist with the CNO cycle of hydrogen at high temperature and are therefore frequently studied e.g. to characterize the different generations of stellar populations of globular clusters \citep{2018ARA&A..56...83B}.     

Our sample of stars is much too small to draw conclusions concerning the success or failure of GCE models. However, the NLTE effects we have demonstrated in this paper are systematic and have a similar effect on stars with similar stellar parameters. It is therefore of interest to compare models and observations to gauge the possible impact for larger samples. Fig.\,\ref{fig:gce} shows how the abundances compare to the GCE model of \citet{2020ApJ...900..179K}.

As seen in Fig.\,\ref{fig:gce}, the mean [Mg/Fe] abundance for metal-poor stars is $0.27$, based on Mg\,I lines (NLTE) and $0.46$, based on Mg\,II lines (LTE). Interestingly, the latter is in significantly better agreement with the plateau value of the model. \citet{2020ApJ...900..179K} report on a similar offset using NLTE literature studies of neutral Mg lines. On the other hand, in the metal-poor regime the GCE model is in excess of the NLTE abundances for neutral lines by $\sim0.2$\,dex for all three elements studied here, which may point to a different explanation, such as a change in the relative abundance of alpha elements due to a change in nuclear reaction rates and/or stellar mass loss \citep[due to rotation and/or binary interaction, see][]{2006ApJ...653.1145K}. We further note that the predicted [(Na,Al)/Fe] ratios could become even higher for GCE models that are based on yields taking into account the effects of stellar rotation.

With such few observational data points as shown in Fig.\,\ref{fig:gce}, these are merely speculations, and we call for a larger investigation of the Mg (and Al) ionization balance to be performed over a large range of metallicities. This is particularly relevant and urgent because these abundances are frequently used to trace the origin of halo stars, and differentiate between ones that formed in-situ at high values compatible with the Galactic thick disk and low values that are interpreted as sign of accretion from lower-mass systems \citep[e.g.][]{2018A&ARv..26....6N,2020ARA&A..58..205H}.    

\section{Conclusions}
\label{sec:conc}
There is a great need to understand how departures from LTE affect chemical abundance determination for all absorption lines that are detectable in stellar spectra. In this work we have computed NLTE abundance corrections with PySME for $2-5$ times more lines for Na, Mg, and Al, compared to what is available in the literature. We show that NLTE analysis generally leads to improved consistency between abundance diagnostics for five benchmark stars and that our NLTE corrections are compatible with previous studies that used the same atomic and atmospheric models, except for strongly saturated lines for which NLTE corrections are very sensitive to the choice of radiative transfer code.

Our conclusions for the different elements are:

\begin{itemize}
    \item Na\,I NLTE corrections are increasingly negative with line strength until the point of maximum saturation. Unsaturated optical lines nearly always have moderately negative corrections. NLTE abundances for different lines are significantly more homogeneous compared to LTE for metal-poor stars. 
    \item Mg\,I NLTE corrections are mostly small in absolute value and can be both positive and negative. The ionization balance is met in the Sun, assuming Mg\,II lines form in LTE, but not in metal-poor unevolved stars. The 0.2\,dex NLTE ionization imbalance for HD84937 and HD140283 can depend on shortcomings in the methodology or the atomic or atmospheric model and must be further investigated. Mg is an important electron donor, and indirect effects on the continuous opacity may lead to unphysical growth curves when the atmospheric model is not consistently updated. Caution is advised when using our data for $\rm[Mg/Fe]<-0.5$ and $\rm[Mg/Fe]>0.5$. 
    \item Al\,I NLTE corrections for metal-rich stars are increasingly negative with line strength until the point of maximum saturation. Unsaturated, optical lines display trends of sharply increasing positive corrections towards lower metallicity for turn-off stars and giants. NLTE analysis strongly improves the ionization balance in the metal-poor regime.
\end{itemize}

The grids of LTE and NLTE equivalent widths are available on the CDS.

\begin{acknowledgements}
      KL and MM acknowledges funds from the European Research Council (ERC) under the European Union’s Horizon 2020 research and innovation programme (Grant agreement No. 852977). The computations were enabled by resources provided by the Swedish National Infrastructure for Computing (SNIC) at UPPMAX, partially funded by the Swedish Research Council through grant agreement no. 2018-05973. CK acknowledges funding from the UK Science and Technology Facility Council (STFC) through grant ST/M000958/1 and ST/V000632/1. This work is supported by the Swedish Research Council through an individual project grant with contract No. 2020-03404. KL and PB acknowledge support by the Knut and Alice Wallenberg Foundation.
\end{acknowledgements}

%
%

\bibliographystyle{aa}
\bibliography{NLTE}

\begin{appendix}

\section{Line data}
This appendix contains four tables, one for each element Na, Mg, Al and Fe, containing line atomic data as well as equivalent widths and abundances for the benchmark stars.  

Specifically, the tables lists for each spectral line the species, the air wavelength in $\AA$\,, the lower level excitation energy in eV, the $\log(gf)$-value, and the Van der Waals broadening parameter (VdW) that represent the logarithm of FWHM per unit pertuber number density at $10,000$\,K, given in $\rm [rad\,s^{-1}\,cm^{-3}]$. If no data is present in this column, the Uns\"{o}ld broadening recipe was used. For each line, there are two or three lines with stellar data that should be interpreted as equivalent width in $\rm m\AA\,$ (first row), the LTE abundance (second row), and NLTE abundance (third row, not available for Mg\,II lines and Fe\,II lines).   

\clearpage
\onecolumn
\begin{longtable}{lccccccccccc}
\caption{\label{table:Na} Na line list}\\
\hline\hline 
& & & & & & \small{Sun}    & \small{Arcturus} & \small{HD84937} & \small{HD140283} & \small{HD122563} \\ 
Ion & $\lambda$ $\rm [\AA\,]$ & Nr. & $E_{\rm low}$ & $\log{gf}$ & VdW &  $W_{\lambda} [\rm m\AA ]$ & $W_{\lambda} [\rm m\AA ]$ & $W_{\lambda} [\rm m\AA ]$  & $W_{\lambda} [\rm m\AA ]$ & $W_{\lambda} [\rm m\AA ]$ \\
 & & & & & & $A\rm(Na)_{LTE}$ & $A\rm(Na)_{LTE}$ & $A\rm(Na)_{LTE}$ & $A\rm(Na)_{LTE}$ & $A\rm(Na)_{LTE}$  \\
 & & & & & & $A\rm(Na)_{NLTE}$ & $A\rm(Na)_{NLTE}$ & $A\rm(Na)_{NLTE}$ & $A\rm(Na)_{NLTE}$ & $A\rm(Na)_{NLTE}$  \\
\hline
\endfirsthead
\caption{continued.}\\
\hline\hline
& & & & & & \small{Sun}    & \small{Arcturus} & \small{HD84937} & \small{HD140283} & \small{HD122563} \\ 
Ion & $\lambda$ $\rm [\AA\,]$ & Nr. & $E_{\rm low}$ & $\log{gf}$ & VdW &  $W_{\lambda} [\rm m\AA ]$ & $W_{\lambda} [\rm m\AA ]$ & $W_{\lambda} [\rm m\AA ]$  & $W_{\lambda} [\rm m\AA ]$ & $W_{\lambda} [\rm m\AA ]$ \\
 & & & & & & $A\rm(Na)_{LTE}$ & $A\rm(Na)_{LTE}$ & $A\rm(Na)_{LTE}$ & $A\rm(Na)_{LTE}$ & $A\rm(Na)_{LTE}$  \\
 & & & & & & $A\rm(Na)_{NLTE}$ & $A\rm(Na)_{NLTE}$ & $A\rm(Na)_{NLTE}$ & $A\rm(Na)_{NLTE}$ & $A\rm(Na)_{NLTE}$  \\
\hline
\endhead
\hline
\endfoot
Na\,I &  3302.368& 6&  0.000&  -1.75$^{(1)}$ &  ...           &   ...         &   ...         &   4.3$\pm$ 0.2&   3.0$\pm$ 0.2&   ...         \\
 & & & & & &                                                      ...           & ...           & 3.99$\pm$ 0.02& 3.40$\pm$ 0.03& ...           \\
 & & & & & &                                                      ...           & ...           & 3.88$\pm$ 0.02& 3.33$\pm$ 0.03& ...           \\
Na\,I &  4497.657&18&  2.104&  -1.53$^{(1)}$ &  ...           &   ...         &   ...         &   ...         &   ...         &   ...         \\
Na\,I &  4668.558&18&  2.104&  -1.27$^{(1)}$ &  ...           &   ...         &  90.6$\pm$ 7.4&   ...         &   ...         &   ...         \\
 & & & & & &                                                      ...           & 6.09$\pm$ 0.13& ...           & ...           & ...           \\
 & & & & & &                                                      ...           & 6.00$\pm$ 0.13& ...           & ...           & ...           \\
Na\,I &  4982.811&18&  2.104&  -0.92$^{(1)}$ &  ...           &  80.0$\pm$ 5.1& 116.5$\pm$ 3.7&   ...         &   ...         &   ...         \\
\end{longtable}
\begin{longtable}{lccccccccccc}
\caption{\label{table:Mg} Mg line list}\\
\hline\hline 
& & & & & & \small{Sun}    & \small{Arcturus} & \small{HD84937} & \small{HD140283} & \small{HD122563} \\ 
Ion & $\lambda$ $\rm [\AA\,]$ & Nr. & $E_{\rm low}$ & $\log{gf}$ & VdW &  $W_{\lambda} [\rm m\AA ]$ & $W_{\lambda} [\rm m\AA ]$ & $W_{\lambda} [\rm m\AA ]$  & $W_{\lambda} [\rm m\AA ]$ & $W_{\lambda} [\rm m\AA ]$ \\
 & & & & & & $A\rm(Mg)_{LTE}$ & $A\rm(Mg)_{LTE}$ & $A\rm(Mg)_{LTE}$ & $A\rm(Mg)_{LTE}$ & $A\rm(Mg)_{LTE}$  \\
 & & & & & & $A\rm(Mg)_{NLTE}$ & $A\rm(Mg)_{NLTE}$ & $A\rm(Mg)_{NLTE}$ & $A\rm(Mg)_{NLTE}$ & $A\rm(Mg)_{NLTE}$  \\
\hline
\endfirsthead
\caption{continued.}\\
\hline\hline
& & & & & & \small{Sun}    & \small{Arcturus} & \small{HD84937} & \small{HD140283} & \small{HD122563} \\ 
Ion & $\lambda$ $\rm [\AA\,]$ & Nr. & $E_{\rm low}$ & $\log{gf}$ & VdW &  $W_{\lambda} [\rm m\AA ]$ & $W_{\lambda} [\rm m\AA ]$ & $W_{\lambda} [\rm m\AA ]$  & $W_{\lambda} [\rm m\AA ]$ & $W_{\lambda} [\rm m\AA ]$ \\
 & & & & & & $A\rm(Mg)_{LTE}$ & $A\rm(Mg)_{LTE}$ & $A\rm(Mg)_{LTE}$ & $A\rm(Mg)_{LTE}$ & $A\rm(Mg)_{LTE}$  \\
 & & & & & & $A\rm(Mg)_{NLTE}$ & $A\rm(Mg)_{NLTE}$ & $A\rm(Mg)_{NLTE}$ & $A\rm(Mg)_{NLTE}$ & $A\rm(Mg)_{NLTE}$  \\
\hline
\endhead
\hline
\endfoot
Mg\,I &  2025.824& 1&  0.000&  -0.98$^{(1)}$ &  ...           &   ...         &   ...         &   ...         &   ...         &   ...         \\
Mg\,I &  2531.458& 2&  2.712&  -1.51$^{(2)}$ & -6.150$^{(2)}$ &   ...         &   ...         &   ...         &   ...         &   ...         \\
Mg\,I &  2546.302& 3&  2.717&  -0.88$^{(2)}$ & -6.310$^{(2)}$ &   ...         &   ...         &   ...         &   ...         &   ...         \\
Mg\,I &  2554.565& 2&  2.712&  -1.16$^{(3)}$ &  ...           &   ...         &   ...         &   ...         &   ...         &   ...         \\
Mg\,I &  2564.936& 3&  2.717&  -0.87$^{(3)}$ &  ...           &   ...         &   ...         &   ...         &   ...         &   ...         \\
Mg\,I &  2776.638& 2&  2.709&  -0.12$^{(4)}$ &  ...           &   ...         &   ...         &  97.5$\pm$ 2.5&  96.0$\pm$ 1.7&   ...         \\
 & & & & & &                                                      ...           & ...           & 5.12$\pm$ 0.03& 4.74$\pm$ 0.02& ...           \\
 & & & & & &                                                      ...           & ...           & 5.29$\pm$ 0.03& 4.96$\pm$ 0.02& ...           \\
\end{longtable}
\begin{longtable}{lccccccccccc}
\caption{\label{table:Al} Al line list}\\
\hline\hline 
& & & & & & \small{Sun}    & \small{Arcturus} & \small{HD84937} & \small{HD140283} & \small{HD122563} \\ 
Ion & $\lambda$ $\rm [\AA\,]$ & Nr. & $E_{\rm low}$ & $\log{gf}$ & VdW &  $W_{\lambda} [\rm m\AA ]$ & $W_{\lambda} [\rm m\AA ]$ & $W_{\lambda} [\rm m\AA ]$  & $W_{\lambda} [\rm m\AA ]$ & $W_{\lambda} [\rm m\AA ]$ \\
 & & & & & & $A\rm(Al)_{LTE}$ & $A\rm(Al)_{LTE}$ & $A\rm(Al)_{LTE}$ & $A\rm(Al)_{LTE}$ & $A\rm(Al)_{LTE}$  \\
 & & & & & & $A\rm(Al)_{NLTE}$ & $A\rm(Al)_{NLTE}$ & $A\rm(Al)_{NLTE}$ & $A\rm(Al)_{NLTE}$ & $A\rm(Al)_{NLTE}$  \\
\hline
\endfirsthead
\caption{continued.}\\
\hline\hline
& & & & & & \small{Sun}    & \small{Arcturus} & \small{HD84937} & \small{HD140283} & \small{HD122563} \\ 
Ion & $\lambda$ $\rm [\AA\,]$ & Nr. & $E_{\rm low}$ & $\log{gf}$ & VdW &  $W_{\lambda} [\rm m\AA ]$ & $W_{\lambda} [\rm m\AA ]$ & $W_{\lambda} [\rm m\AA ]$  & $W_{\lambda} [\rm m\AA ]$ & $W_{\lambda} [\rm m\AA ]$ \\
 & & & & & & $A\rm(Al)_{LTE}$ & $A\rm(Al)_{LTE}$ & $A\rm(Al)_{LTE}$ & $A\rm(Al)_{LTE}$ & $A\rm(Al)_{LTE}$  \\
 & & & & & & $A\rm(Al)_{NLTE}$ & $A\rm(Al)_{NLTE}$ & $A\rm(Al)_{NLTE}$ & $A\rm(Al)_{NLTE}$ & $A\rm(Al)_{NLTE}$  \\
\hline
\endhead
\hline
\endfoot
Al\,I &  2118.341& 1&  0.000&  -1.56$^{(1)}$ &  ...           &   ...         &   ...         &   ...         &   8.6$\pm$ 1.0&   ...         \\
 & & & & & &                                                      ...           & ...           & ...           & 3.16$\pm$ 0.06& ...           \\
 & & & & & &                                                      ...           & ...           & ...           & 3.49$\pm$ 0.05& ...           \\
Al\,I &  2134.747& 2&  0.014&  -1.10$^{(1)}$ &  ...           &   ...         &   ...         &   ...         &   ...         &   ...         \\
Al\,I &  2145.555& 1&  0.000&  -1.25$^{(1)}$ &  ...           &   ...         &   ...         &   ...         &  17.8$\pm$ 0.4&   ...         \\
 & & & & & &                                                      ...           & ...           & ...           & 3.23$\pm$ 0.01& ...           \\
 & & & & & &                                                      ...           & ...           & ...           & 3.57$\pm$ 0.01& ...           \\
Al\,I &  2204.626& 7&  0.000&  -0.90$^{(1)}$ &  ...           &   ...         &   ...         &   ...         &  36.5$\pm$ 0.8&   ...         \\
\end{longtable}
\begin{longtable}{lccccccccccc}
\caption{\label{table:Fe} Fe line list}\\
\hline\hline 
& & & & & & \small{Sun}    & \small{Arcturus} & \small{HD84937} & \small{HD140283} & \small{HD122563} \\ 
Ion & $\lambda$ $\rm [\AA\,]$ & Nr. & $E_{\rm low}$ & $\log{gf}$ & VdW &  $W_{\lambda} [\rm m\AA ]$ & $W_{\lambda} [\rm m\AA ]$ & $W_{\lambda} [\rm m\AA ]$  & $W_{\lambda} [\rm m\AA ]$ & $W_{\lambda} [\rm m\AA ]$ \\
 & & & & & & $A\rm(Fe)_{LTE}$ & $A\rm(Fe)_{LTE}$ & $A\rm(Fe)_{LTE}$ & $A\rm(Fe)_{LTE}$ & $A\rm(Fe)_{LTE}$  \\
\hline
\endfirsthead
\caption{continued.}\\
\hline\hline
& & & & & & \small{Sun}    & \small{Arcturus} & \small{HD84937} & \small{HD140283} & \small{HD122563} \\ 
Ion & $\lambda$ $\rm [\AA\,]$ & Nr. & $E_{\rm low}$ & $\log{gf}$ & VdW &  $W_{\lambda} [\rm m\AA ]$ & $W_{\lambda} [\rm m\AA ]$ & $W_{\lambda} [\rm m\AA ]$  & $W_{\lambda} [\rm m\AA ]$ & $W_{\lambda} [\rm m\AA ]$ \\
 & & & & & & $A\rm(Fe)_{LTE}$ & $A\rm(Fe)_{LTE}$ & $A\rm(Fe)_{LTE}$ & $A\rm(Fe)_{LTE}$ & $A\rm(Fe)_{LTE}$  \\
\hline
\endhead
\hline
\endfoot
Fe\,II&  4923.921& 1&  2.891&  -1.26$^{(1)}$ & -7.884$^{(3)}$ &   ...         &   ...         &  63.2$\pm$ 0.4&  58.6$\pm$ 0.4& 104.4$\pm$ 0.6\\
 & & & & & &                                                      ...           & ...           &-2.08$\pm$ 0.01&-2.49$\pm$ 0.01&-2.52$\pm$ 0.01\\
Fe\,II&  4993.350& 1&  2.807&  -3.68$^{(2)}$ & -7.894$^{(3)}$ &  39.8$\pm$ 0.4&   ...         &   1.8$\pm$ 0.1&   2.1$\pm$ 0.1&   6.4$\pm$ 0.1\\
 & & & & & &                                                      0.01$\pm$ 0.01& ...           &-1.96$\pm$ 0.03&-2.13$\pm$ 0.03&-2.50$\pm$ 0.01\\
Fe\,II&  5169.028& 1&  2.891&  -1.00$^{(1)}$ & -7.887$^{(3)}$ &   ...         &   ...         &  76.9$\pm$ 1.6&  72.9$\pm$ 2.7&   ...         \\
 & & & & & &                                                      ...           & ...           &-2.02$\pm$ 0.03&-2.43$\pm$ 0.07& ...           \\
Fe\,II&  5197.567& 1&  3.230&  -2.22$^{(1)}$ & -7.880$^{(3)}$ &  83.6$\pm$ 0.6&   ...         &  13.0$\pm$ 0.2&  10.8$\pm$ 0.1&  39.0$\pm$ 0.3\\
 & & & & & &                                                     -0.11$\pm$ 0.02& ...           &-2.12$\pm$ 0.01&-2.43$\pm$ 0.01&-2.49$\pm$ 0.01\\
\end{longtable}
\end{appendix}

\end{document}